\documentclass[aps,prd,twocolumn,groupedaddress,showpacs,nofootinbib]{revtex4}

\usepackage{graphicx}

\begin{document}

\preprint{UTAP-439}
\preprint{RESCEU-12/03}

\title{Resonant spin-flavor conversion of supernova neutrinos:
Dependence on presupernova models and future prospects}

\author{Shin'ichiro Ando}
\email[Email address: ]{ando@utap.phys.s.u-tokyo.ac.jp}
\affiliation{Department of Physics, School of Science, the University of
Tokyo, 7-3-1 Hongo, Bunkyo-ku, Tokyo 113-0033, Japan}
\author{Katsuhiko Sato}
\affiliation{Department of Physics, School of Science, the University of
Tokyo, 7-3-1 Hongo, Bunkyo-ku, Tokyo 113-0033, Japan}
\affiliation{Research Center for the Early Universe, School of Science,
the University of Tokyo, 7-3-1 Hongo, Bunkyo-ku, Tokyo 113-0033, Japan}

\date{Received 22 March 2003; accepted 3 May 2003}

\begin{abstract}

We study the resonant spin-flavor (RSF) conversion of supernova
 neutrinos, which is induced by the interaction between the nonzero
 neutrino magnetic moment and the supernova magnetic fields, and its
 dependence on presupernova models.
As the presupernova models, we adopt the latest ones by Woosley, Heger,
 and Weaver, and, further, models with both solar and zero metallicity
 are investigated.
Since the $(1-2Y_e)$ profile of the new presupernova models, which is
 responsible for the RSF conversion, suddenly drops at the resonance
 region, the completely adiabatic RSF conversion is not realized, even
 if $\mu_\nu B_0= (10^{-12}\mu_B)(10^{10}~{\rm G})$, where $B_0$ is the
 strength of the magnetic field at the surface of the iron core.
In particular for the model with zero metallicity, the conversion is
 highly nonadiabatic in the high energy region, reflecting the
 $(1-2Y_e)$ profile of the model.
In calculating the flavor conversion, we find that the shock wave
 propagation, which changes density profiles drastically, is a much more
 severe problem than it is for the pure Mikheyev-Smirnov-Wolfenstein
 (MSW) conversion case.
This is because the RSF effect occurs at a far deeper region than the
 MSW effect.
To avoid the uncertainty concerning the shock propagation, we restrict
 our discussion to 0.5 s after the core bounce (and for more
 conservative discussion, 0.25 s), during which the shock wave is not
 expected to affect the RSF region.
We also evaluate the energy spectrum at the Super-Kamiokande detector
 for various models using the calculated conversion probabilities, and
 find that it is very difficult to obtain useful information on the
 supernova metallicities and magnetic fields or on the neutrino magnetic
 moment from the supernova neutrino observation.
Future prospects are also discussed.
\end{abstract}


\maketitle

\section{INTRODUCTION \label{sec:INTRODUCTION}}

A core-collapse supernova explosion is one of the most spectacular
events in astrophysics; 99\% of its gravitational binding energy is
released as neutrinos, and only 1\% as the kinetic energy of the
explosion.
Therefore, neutrinos play an essential role in supernovae, and their
detection by ground-based large water \v{C}erenkov detectors, such as
Super-Kamiokande (SK) and Sudbury Neutrino Observatory (SNO), would
provide valuable information not only on supernova physics but also on
the nature of neutrinos.
What we can learn from the next galactic supernova has been considered
in many articles (for a review, see Ref. \citep{Raffelt02}).
For example, we can constrain the properties of neutrino oscillations,
such as the mixing angle between the first and third mass eigenstates
($\theta_{13}$), and the mass hierarchy [normal ($m_1\ll m_3$) or
inverted ($m_1\gg m_3$)] \citep{Dighe00,Takahashi02b}.

In addition to the nonzero neutrino masses and mixing angles, the
nonzero magnetic moment is of a different nature for neutrinos beyond the
standard model of particle physics, and has attracted a great deal of
attention from many theoretical and experimental physicists.
If neutrinos have a nonzero magnetic moment, it leads to precession
between left- and right-handed neutrinos in sufficiently strong magnetic
fields \citep{Cisneros70,Fujikawa80}.
In general, nondiagonal elements of the magnetic moment matrix are
possible and neutrinos can be changed into different flavors and
chiralities \citep{Schechter81,Schechter82}.
Furthermore, with the additional effect of coherent forward scattering
by matter, neutrinos can be resonantly converted into those with
different chiralities \citep{Lim88,Akhmedov88a,Akhmedov88b} by a
mechanism similar to the well-known Mikheyev-Smirnov-Wolfenstein (MSW)
effect \citep{Wolfenstein78,Mikheyev85,Mikheyev86}.
This resonant spin-flavor (RSF) conversion induced by the neutrino
magnetic moment in strong magnetic fields was first introduced to
solve the solar neutrino problem, and actually gave the best fit
solution before the KamLAND result \citep{Barranco02}.
However, the recent KamLAND experiment \citep{Eguchi03} has shown that
the large mixing angle MSW solution is the most favorable one; the RSF
mechanism is suppressed at the subdominant level.
From the KamLAND negative results for the solar antineutrino search, an
upper bound on the neutrino magnetic moment is obtained, $\mu_\nu\alt 1
\times 10^{-12}\mu_B$, where $\mu_B$ is the Bohr magneton
\citep{Torrente-Lujan03}.
This upper bound is comparable to the most stringent limit from the
stellar cooling argument, $\mu_\nu\alt$(1--4)$\times 10^{-12}\mu_B$
\citep{Ayala99}.

Although the RSF mechanism does not work at a dominant level in the
Sun, it may occur efficiently in a denser environment with stronger
magnetic field, which is actually expected in the case of core-collapse
supernovae.
The RSF conversion mechanism in supernovae has been investigated by many
authors \citep{Lim88,Akhmedov88a,Akhmedov88b,Voloshin88,Akhmedov92,
Akhmedov93a,Peltoniemi92,Athar95,Totani96b,Nunokawa97,Nunokawa99,
Ando03b}.
Among them, \citet{Ando03b} studied the RSF effect using a three-flavor
formulation with the latest oscillation parameters, and pointed out that
the combination of the MSW and RSF effects makes the crossing scheme
very interesting to investigate.
They showed that until 0.5 s after the core bounce the RSF-induced
$\bar\nu_e\leftrightarrow\nu_\tau$ transition occurs efficiently, when
$\mu_\nu\agt 10^{-12}\mu_B(B_0/5\times 10^9~{\rm G})^{-1}$, where $B_0$
is the strength of the magnetic field at the surface of the iron core.

The effective matter potential for the RSF conversion is given in a
form proportional to the value of $(1-2Y_e)$, where $Y_e$ is the
electron number fraction per nucleon.
Thus, the deviation of the value of $Y_e$ from 0.5 in the stellar
envelope is quite important, and this value is strongly dependent on the
isotopic composition.
Since this deviation is determined by rarely existent nuclei, an
accurate estimate of this deviation is quite difficult.
Therefore, the astrophysical uncertainty in $(1-2Y_e)$ should be
discussed.
This point was first investigated by \citet{Totani96b}.
However, their treatment was based on a two-flavor formulation with
the large uncertainties concerning the mixing parameters of those days.
Therefore, we need definitive investigations using the three-flavor
formulation with the latest presupernova models as well as the recently
determined neutrino parameters.

In this paper, we study the RSF conversion mechanism using the
three-flavor formulation with the latest neutrino mixing parameters.
In particular, we investigate the dependence on presupernova models; we
use the latest $15M_\odot$ model by \citet{Woosley02},
and compare the results with those obtained with the previous progenitor
model by \citet{Woosley95}, which was also adopted in the calculations
of \citet{Ando03b}.
It is also expected that the value of $(1-2Y_e)$ strongly depends on the
stellar metallicity, and hence we use the $15M_\odot$ model with two
different metallicities, solar and zero metallicity, and the metallicity
effect is investigated.

Throughout this paper, we adopt the realistic neutrino mixing parameters
inferred from the recent experimental results: for the atmospheric
neutrino parameters, $\Delta m_{13}^2=2.8\times 10^{-3}~{\rm eV}^2,
\sin^22\theta_{23}=1.0$, and for the solar neutrino parameters, $\Delta
m_{12}^2=5.0\times 10^{-5}~{\rm eV}^2,\tan^2\theta_{12}=0.42$.
As for the still uncertain parameter $\theta_{13}$, we assume $\sin^22
\theta_{13}=10^{-6}$,\footnote{As already shown in Ref. \citep{Ando03b},
the dependence of the detected event on the parameter $\theta_{13}$ is
very weak, and even if we assume a large value of $\theta_{13}$, the
results do not change.} and we also assume a normal mass hierarchy.
For the elements of the neutrino magnetic moment matrix $\mu_{ij}$,
where $i$ and $j$ denote the flavor eigenstates of the neutrinos, i.e.,
$e,\mu,$ and $\tau$, we assume that all the values of $\mu_{ij}$ are
near the current upper limit, or $\mu_{ij}=10^{-12}\mu_B$.

The remainder of this paper is organized as follows.
In Sec. \ref{sec:FORMULATION AND LEVEL CROSSING SCHEME}, we give the
formulation used in our calculation, which includes all three-flavor
neutrinos and antineutrinos, and from the formulation a level crossing
diagram, which enables us to understand the conversion scheme
intuitively, is introduced.
In Sec. \ref{sec:PRESUPERNOVA MODELS}, the presupernova model we adopt
in our calculations is illustrated, and the results of numerical
calculations are shown in Sec. \ref{sec:RESULTS}.
Finally, detailed discussions of the model uncertainties and of whether
we can obtain, at present or in the future, useful information on
physical and astrophysical quantities are presented in
Sec. \ref{sec:DISCUSSION}.

\section{FORMULATION AND LEVEL CROSSING SCHEME \label{sec:FORMULATION
AND LEVEL CROSSING SCHEME}}

\subsection{Interaction with matter and magnetic fields
\label{sub:Interaction with matter and magnetic fields}}

The interaction of the magnetic moment of neutrinos and magnetic fields
is described by
\begin{equation}
\langle(\nu_i)_R|H_{\mathrm{int}}|(\nu_j)_L \rangle  = \mu_{ij} B_\perp ,
\label{eq:interaction Hamiltonian}
\end{equation}
where $B_\perp$ is the magnetic field transverse to the direction of
propagation, and $(\nu)_R$ and $(\nu)_L$ are the right- and left-handed
neutrinos, respectively.
If neutrinos are Dirac particles, right-handed neutrinos and
left-handed antineutrinos are undetectable (sterile neutrinos), since
they do not interact with matter.
On the other hand, if neutrinos are Majorana particles, $\nu_R$'s
are identical to antiparticles of $\nu_L$'s and interact with matter.
In this paper, we assume that neutrinos are Majorana particles.
The diagonal magnetic moments are forbidden for Majorana neutrinos,
and therefore only conversion between different flavors is
possible, e.g., $(\bar{\nu}_e)_R \leftrightarrow (\nu_{\mu,\tau})_L$.

Coherent forward scattering with matter induces an effective
potential for neutrinos, which is calculated using weak interaction
theory.
The effective potential due to scattering with electrons is given by
\begin{equation}
V_{\pm\pm} = \pm \sqrt{2} G_F \left( \pm \frac{1}{2} 
			       + 2\sin^2 \theta_W \right) n_e, 
\label{eq:electron potential}
\end{equation}
where $n_e$ is the electron number density, $G_F$ is the Fermi coupling
constant, and $\theta_W$ is the Weinberg angle.
The $\pm$ sign in front refers to $\nu~(+)$ and $\bar{\nu}~(-)$ and that
in the parentheses to $\nu_e~(+)$ and $\nu_{\mu,\tau}~(-)$.
The difference between $e$ and $\mu,\tau$ neutrinos comes from the
existence of charged-current interaction.
The subscript $\pm\pm$ of $V$ refers to the first and the second $\pm$
sign.
The ordinary MSW effect between $\nu_e$ and $\nu_{\mu,\tau}$ is caused by
the potential difference $V_e-V_{\mu,\tau} = V_{++}-V_{+-} = \sqrt{2}
G_F n_e$.
To include the RSF effect, which causes conversion between neutrinos
and antineutrinos, we should take into account the neutral-current
scattering by nucleons:
\begin{equation}
V = \sqrt{2} G_F \left(\frac{1}{2} - 2 \sin^2 \theta_W \right) n_p 
 - \sqrt{2} G_F \frac{1}{2} n_n,
\label{eq:nucleon potential}
\end{equation}
where $n_p,n_n$ are the proton and neutron number density,
respectively.
For neutrinos we add $+V$ to the potential and for antineutrinos $-V$.
Therefore, the RSF conversion between $\bar{\nu}_e$ and $\nu_{\mu,\tau}$
obeys the potential difference 
\begin{eqnarray}
\Delta V &\equiv& V_{\bar{e}}-V_{\mu,\tau} \nonumber \\
 &=& ( V_{-+} - V )  - ( V_{+-} + V )  \nonumber \\
 &=& \sqrt{2} G_F \frac{\rho}{m_N} ( 1 - 2 Y_e ),
\label{eq:Delta V}
\end{eqnarray}
where $\rho$ is the density, $m_N$ is the nucleon mass, and $Y_e=n_e/
(n_e+n_n)$ is the number of electrons per baryon. [When we obtained
Eq. (\ref{eq:Delta V}), we assumed charge neutrality $n_e=n_p$.]

\subsection{Three-flavor formulation \label{sub:Three-flavor
formulation}}

The time evolution of the flavor mixing, which includes all three-flavor
neutrinos and antineutrinos, is described by the Schr\"odinger equation
\begin{equation}
i \frac{d}{dr} \left(
	      \begin{array}{c} 
	       \nu \\
	       \bar{\nu}
	      \end{array}
	      \right)
= \left( \begin{array}{cc}
 H_0 & B_\perp M \\ -B_\perp M & \bar H_0 \end{array} \right)
\left( \begin{array}{c} \nu \\ \bar{\nu} \end{array} \right),
\label{eq:three-flavor}
\end{equation}
where 
\begin{widetext}
\begin{equation}
\nu = \left(
	 \begin{array}{c}
	  \nu_e \\
	  \nu_\mu \\
	  \nu_\tau
	 \end{array}
       \right), 
~~\bar{\nu} = \left(
	 \begin{array}{c}
	  \bar{\nu}_e \\
	  \bar{\nu}_\mu \\
	  \bar{\nu}_\tau
	 \end{array}
       \right), 
\end{equation}
 \begin{eqnarray}
  H_0 = \frac{1}{2E_\nu} U \left(
			  \begin{array}{ccc}
			  0 & 0 & 0 \\
			  0 & \Delta m^2_{12} & 0 \\
			  0 & 0 & \Delta m^2_{13}
			  \end{array}
			  \right) U^\dagger
  + \left(
   \begin{array}{ccc}
   V_{++}+V & 0 & 0 \\
   0 & V_{+-}+V & 0 \\
   0 & 0 & V_{+-}+V 
   \end{array}
   \right), \label{equationa} 
\\
  \bar H_0 = \frac{1}{2E_\nu} U \left(
			  \begin{array}{ccc}
			  0 & 0 & 0 \\
			  0 & \Delta m^2_{12} & 0 \\
			  0 & 0 & \Delta m^2_{13}
			  \end{array}
			  \right) U^\dagger
  + \left(
   \begin{array}{ccc}
   V_{-+}-V & 0 & 0 \\
   0 & V_{--}-V & 0 \\
   0 & 0 & V_{--}-V 
   \end{array}
   \right), \label{equationb}
\\
  U = \left(
       \begin{array}{ccc}
	U_{e1} & U_{e2} & U_{e3} \\
        U_{\mu 1} & U_{\mu 2} & U_{\mu 3} \\
        U_{\tau 1} & U_{\tau 2} & U_{\tau 3}
       \end{array}
     \right)
  = \left(
   \begin{array}{ccc}
   c_{12}c_{13} 
    & s_{12}c_{13} 
   & s_{13} \\
   -s_{12}c_{23}
    -c_{12}s_{23}s_{13}
  & c_{12}c_{23}
  -s_{12}s_{23}s_{13} 
  & s_{23}c_{13} \\
    s_{12}s_{23}
     -c_{12}c_{23}s_{13}
     & -c_{12}s_{23}
     -s_{12}c_{23}s_{13}
     & c_{23}c_{13}
     \end{array}
   \right), \label{eq:U}
 \end{eqnarray}
\begin{equation}
M = \left(
   \begin{array}{ccc}
   0 & \mu_{e\mu} & \mu_{e\tau} \\
   -\mu_{e\mu} & 0 & \mu_{\mu\tau} \\
   -\mu_{e\tau} & -\mu_{\mu\tau} & 0
   \end{array}
   \right),
\label{eq:magnetic moment}
\end{equation}
\end{widetext}
and $c_{ij}=\cos\theta_{ij},~s_{ij}=\sin\theta_{ij}$.
[We assume the $CP$ phase $\delta=0$ in Eq. (\ref{eq:U}) for simplicity.]

Resonant flavor conversion occurs when two diagonal elements of the
matrix in Eq. (\ref{eq:three-flavor}) have the same value.
There are five such resonance points, which are for $\nu_e
\leftrightarrow \nu_\mu$ (MSW-L), $\nu_e \leftrightarrow \nu_\tau$
(MSW-H), $\bar{\nu}_e \leftrightarrow \nu_\mu$ (RSF-L), $\bar{\nu}_e
\leftrightarrow \nu_\tau$ (RSF-H), and $\bar{\nu}_\mu \leftrightarrow
\nu_\tau$ conversions.
The suffixes ``-L'' and ``-H'' attached to ``MSW'' and ``RSF'' indicate
whether the density at the resonance points is lower or higher.
Hereafter, we neglect the $\bar{\nu}_\mu \leftrightarrow \nu_\tau$
conversion, since it is always nonadiabatic and including it further
complicates the discussion.

\begin{figure}[htbp]
\begin{center}
\includegraphics[width=8cm]{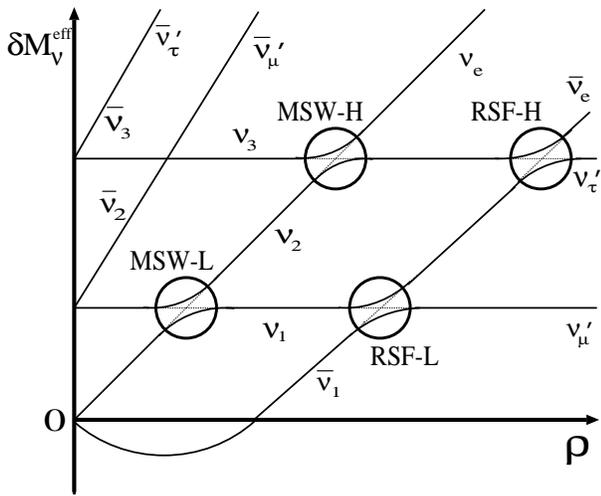}
\caption{Schematic illustration of level crossings, where
 $\nu_{1,2,3}$ and $\bar{\nu}_{1,2,3}$ represent the mass eigenstates of
 neutrinos and antineutrinos in matter, respectively, and
 $\nu_{\mu,\tau}^\prime$ and $\bar{\nu}_{\mu,\tau}^\prime$ the mass
 eigenstates at production, which are superpositions of $\nu_\mu$ and
 $\nu_\tau$ or $\bar{\nu}_\mu$ and $\bar{\nu}_\tau$. There are four
 resonance points, MSW-L, MSW-H, RSF-L, and RSF-H. Adiabatic
 conversion means that the neutrinos trace the solid curve at each
 resonance point (i.e., the mass eigenstate does not flip), while
 nonadiabatic conversion is shown by the dotted
 line. \label{fig:crossing}}
\end{center}
\end{figure}

Figure \ref{fig:crossing} shows the level crossing diagram, which we
introduced in our previous paper \citep{Ando03b}, to understand
the flavor conversions described by Eq. (\ref{eq:three-flavor})
intuitively.
The figure clearly includes not only the ordinary MSW resonances but
also the RSF effects, and it is expected that the combined effect of MSW
and RSF makes this scheme very interesting to investigate.
For instance, when MSW-L and RSF-H are adiabatic and the others are
nonadiabatic (this case is actually expected if $\theta_{13}$ is small
and the magnetic field is sufficiently strong), conversions such as
$\nu_e\to\nu_2,\nu_\mu^\prime\to\nu_1,\nu_\tau^\prime\to\bar\nu_1,
\bar\nu_e\to\nu_3,\bar\nu_\mu^\prime\to\bar\nu_2$, and
$\bar\nu_\tau^\prime\to\bar\nu_3$ occur.
We can easily predict this sort of conversion scheme from
Fig. \ref{fig:crossing}, when all the resonances are either completely
adiabatic or completely nonadiabatic; for the intermediate cases we have
no choice but to trust numerical calculations.

\section{PRESUPERNOVA MODELS \label{sec:PRESUPERNOVA MODELS}}

\subsection{Density and $Y_e$ profiles \label{sub:Density and $Y_e$
profiles}}

In this subsection, we discuss the several presupernova models that we
adopt in our transition calculations.
In a previous paper \citep{Ando03b}, we used only the presupernova
model of \citet{Woosley95}, which is for $15M_\odot$ and solar
metallicity (hereafter W95S, where ``S'' denotes solar metallicity).
However, we should investigate the dependence on adopted presupernova
models with various metallicities.
This is because the RSF conversion is very sensitive to the deviation of
$Y_e$ from 0.5 [see Eq. (\ref{eq:Delta V})], which strongly depends on
the metallicities as well as on the weak interaction rates adopted in
the simulation of stellar evolution.

\begin{figure}[htbp]
\begin{center}
\includegraphics[width=8cm]{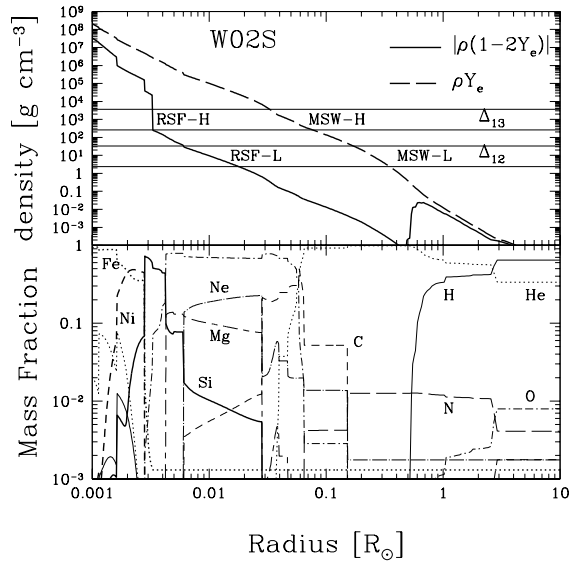}
\caption{Presupernova profiles (W02S) used in our calculations taken
 from Ref. \citep{Woosley02}. Upper panel: The density and $Y_e$
 combination that is responsible for the RSF conversions [$|\rho
 (1-2Y_e)|$, solid curve], and that for the MSW conversions ($\rho Y_e$,
 dashed curve). Two horizontal bands represent $\Delta_{12}$ and
 $\Delta_{13}$ (these definitions are given in the text); at the
 intersections between them and the profile curves, the RSF and MSW
 conversions occur. Lower panel: The mass fraction of the various
 elements. \label{fig:profiles_s15}}
\end{center}
\end{figure}

\begin{figure}[htbp]
\begin{center}
\includegraphics[width=8cm]{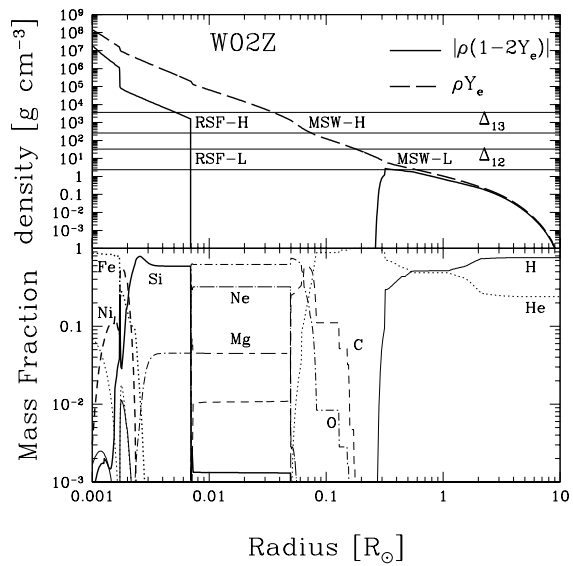}
\caption{The same as Fig. \ref{fig:profiles_s15}, but for the
 W02Z model. \label{fig:profiles_z15}}
\end{center}
\end{figure}

In this study, we adopt the latest presupernova models by
\citet{Woosley02} with both solar and zero metallicities: W02S and
W02Z (``Z'' denotes zero metallicity).
Figures \ref{fig:profiles_s15} and \ref{fig:profiles_z15} show density
profiles (upper panel) or $\rho Y_e$, which is responsible for MSW, and
$|\rho (1-2Y_e)|$ for RSF, and the composition of each element (lower
panel), for the W02S and W02Z models, respectively.
In the upper panels of both figures, we also show $\Delta_{12}\equiv
m_N\Delta m_{12}^2\cos 2\theta_{12}/2\sqrt 2G_FE_\nu$ and $\Delta_{13}
\equiv m_N\Delta m_{13}^2\cos 2\theta_{13}/2\sqrt 2G_FE_\nu$ as two
horizontal bands (the bandwidth comes from the energy range 5--70 MeV).
At intersections between $\Delta_{12},\Delta_{13}$ and $\rho (1-2Y_e),
\rho Y_e$, the RSF and MSW conversions take place.

For the W02S model, the region where RSF-H occurs is the silicon burning
shell, in which $(1-2Y_e)$ suddenly drops, in contrast with RSF-L, where
$\rho(1-2Y_e)$ gradually changes.
On the other hand, for the W02Z model, $(1-2Y_e)$ suddenly becomes
exactly zero at the boundary between Si + O and O + Ne + Mg layers, and
this tendency continues to the He layer because of the lack of heavy
nuclei, which cause the deviation of $Y_e$ from 0.5.
In consequence, both the RSF-H and RSF-L conversions are expected to be
highly nonadiabatic, as discussed in more detail in
Sec. \ref{sec:RESULTS}.
Note that the $\rho Y_e$ profiles of the two different models agree well
with each other, which clearly indicates that the ordinary MSW
conversions are not noticeably affected by the value of $Y_e$.

\begin{figure}[htbp]
\begin{center}
\includegraphics[width=8cm]{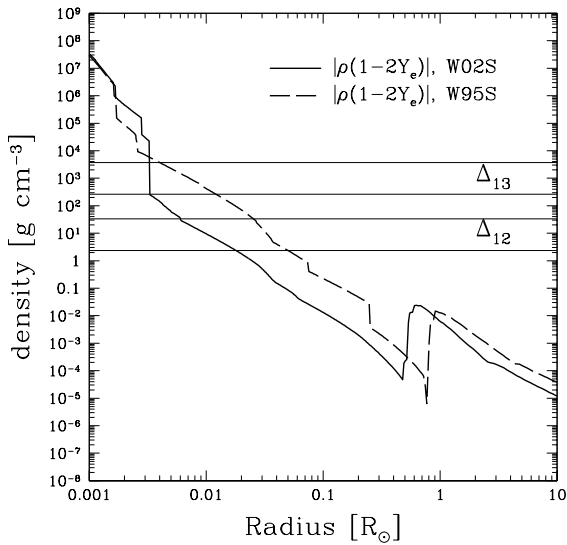}
\caption{The $|\rho (1-2Y_e)|$ profiles, which are responsible for the
 RSF conversions, for the W02S (solid curve) and W95S (dashed curve)
 models. \label{fig:profiles}}
\end{center}
\end{figure}

In Fig. \ref{fig:profiles}, we compare the $|\rho (1-2Y_e)|$ profile of
the W02S model with that of the W95S model, which was adopted in
previous publications including Ref. \citep{Ando03b}.
The difference comes from the included weak interaction rates for
nuclei.
In the latest model W02S, a recent shell model is included in the
calculations and results in substantial revisions to the older data set
in the W95S model (see Ref. \citep{Heger01} for a detailed discussion).
As a result, the value of $(1-2Y_e)$ decreases by a few orders of
magnitude between the He and Si + O shells.
Particularly with the new model W02S, there is a sudden drop of $\rho
(1-2Y_e)$ when the RSF-H conversion occurs, in contrast with the gradual
decrease in the case of the W95S model.
Since the adiabaticity of the resonance is reciprocal to the gradient of
the logarithmic value of $\rho (1-2Y_e)$, the RSF conversion is expected
to be less efficient than that with W95S given in Ref. \citep{Ando03b}.

Although we use static progenitor models in calculating the flavor
transition, in fact the density profile changes drastically during a
neutrino burst ($\sim 10$ s) owing to shock wave propagation, and we
should use the time-dependent profiles \citep{Schirato02}.
Unfortunately, however, supernova explosion mechanisms are still
controversial, and there is no reliable model that precisely describes
the time-dependent density and $Y_e$ profiles.
Further, there is also a large uncertainty concerning the magnetic field
structure which is affected by the shock wave propagation.
Therefore, from this point on, we confine our discussion to 0.5 s after
core bounce, since in that case using the static presupernova and
magnetic field models is considered to be a good approximation
\citep{Ando03b}.
This is based on the numerical calculation by \citet{Takahashi02c}, the
only authors having succeeded in shock propagation to the outer
envelope.
Although we cannot trust the details of their result without any doubt,
the choice of the time scale during which the shock effect can be
neglected is expected to be reasonable.
In addition, for more conservative discussions, we also give
calculations during first 0.25 s.

\subsection{Magnetic fields \label{sub:Magnetic fields}}

We assume that the global structure of the magnetic field is a dipole
moment and the field strength is normalized at the surface of the iron
core with the values of $10^8$ and $10^{10}$ G.
The reason for this normalization is as follows.
The magnetic fields should be normalized by fields that are static
and exist before the core collapse, because those of a nascent
neutron star can hardly affect the far outer region, where the RSF
conversions take place, within the short time scale of a neutrino burst.
As discussed in the previous subsection, since the shock wave does not
affect the resonance region at $\alt 0.5$ s after bounce, it is also
expected that the magnetic field structure and strength at the resonance
points are not seriously changed at that time.
The strength of such magnetic fields above the surface of the iron core
may be inferred from observations of the surface of white dwarfs, since
both are sustained against gravitational collapse by the degenerate
pressure of electrons.
Observations of the magnetic fields in white dwarfs show that the
strength spreads in a wide range of $10^7$--$10^9$ G
\cite{Chanmugam92}.
Considering the possibility of the decay of magnetic fields in white
dwarfs, it is not unnatural to consider magnetic fields up to
$10^{10}$ G at the surface of the iron core.
Then, in Eq. (\ref{eq:three-flavor}), $B_\perp = B_0 (r_0 / r)^3 \sin
\Theta$, where $B_0$ is the strength of the magnetic field at the
equator on the iron core surface, $r_0$ the radius of the iron core, and
$\Theta$ the angle between the pole of the magnetic dipole and the
direction of neutrino propagation. 
Hereafter, we assume $\sin \Theta = 1$.

\section{RESULTS \label{sec:RESULTS}}

\subsection{Conversion probabilities \label{sub:Conversion
probabilities}}

\begin{figure}[htbp]
\begin{center}
\includegraphics[width=8cm]{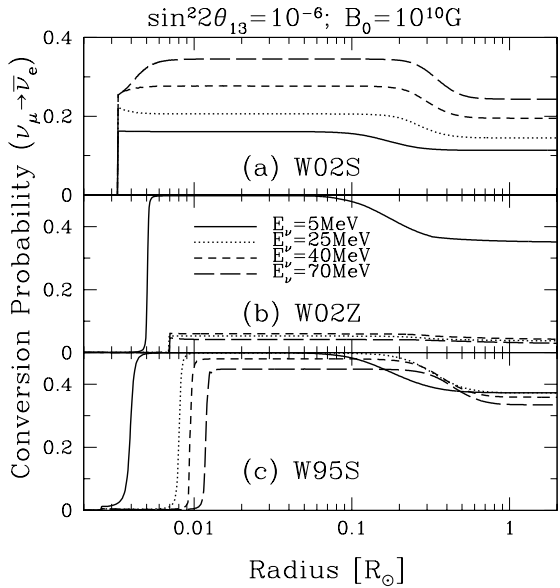}
\caption{Conversion probability $P(\nu_\mu\to\bar\nu_e)$ as a function
 of radius for $B_0=10^{10}$ G. The probabilities calulated with the
 W02S (a), W02Z (b), and W95S (c) models are
 plotted. \label{fig:prob_em}}
\end{center}
\end{figure}

We calculated Eq. (\ref{eq:three-flavor}) numerically with the adopted
models given in Sec. \ref{sec:PRESUPERNOVA MODELS}, and obtained the
conversion probabilities for each flavor.
Among them, we show in Fig. \ref{fig:prob_em} those of the
$\nu_\mu\to\bar\nu_e$ transition for various presupernova models in the
case of $B_0=10^{10}$ G.
(This conversion channel is essential in order to discuss the efficiency
of the RSF effects.)

For the W02S model, Fig. \ref{fig:prob_em}(a) shows that the conversion
between $\nu_{\mu,\tau}$ and $\bar\nu_e$ occurs at a radius
independent of energy, and becomes adiabatic as the energy increases;
this character reflects the $\rho (1-2Y_e)$ profile given in
Fig. \ref{fig:profiles_s15}, which shows a sudden drop in the RSF-H
region.

The behavior changes dramatically when the metallicity is zero, or for
the W02Z model.
In this case, as shown in Fig. \ref{fig:prob_em}(b), the transition is
completely adiabatic at low energy; however, once the energy is
increased beyond some critical value, the conversion abruptly becomes
almost completely nonadiabatic.
This tendency also reflects the characteristic profile in
Fig. \ref{fig:profiles_z15}.

The flavor transition occurs most efficiently in the case of the W95S
model as shown in Fig. \ref{fig:prob_em}(c), which indicates the most
moderate profile at each resonance point (see the dashed curve in
Fig. \ref{fig:profiles}).\footnote{Actually, the results for the W95S
model are taken from previous calculations (Ref. \citep{Ando03b})
with a slightly different mixing parameter, $\tan^2\theta_{12}=0.34$
instead of 0.42 in the new calculations. However, note that the
difference due to the parameter choice is quite small, as shown in the
next subsection.}
With a sufficiently strong magnetic field $B_0=10^{10}$ G, the $\nu_\mu
\to\bar\nu_e$ transition is almost completely adiabatic over the entire
energy range.

For all these three models, it appears that the RSF-L conversion does
not play any role.
This is because for the W02S and W95S models RSF-L occurs farther out
than RSF-H, where the magnetic field strength is not large enough to
induce adiabatic conversions.
On the other hand, for the W02Z model, the RSF-L occurs in the same
place as RSF-H; the magnetic field strength is also the same for these
two resonances.
In this case, however, the very abrupt drop of the $\rho(1-2Y_e)$
profile at RSF-L strongly suppresses efficient flavor conversions.

At the end of this subsection, we focus on the decrease of the
conversion probability at $\agt 0.1R_\odot$ which can be seen in
Fig. \ref{fig:prob_em}.
This does not indicate any resonances but is merely an effect of flavor
mixings.
At such a deep region in the supernova envelope as $\alt 0.1R_\odot$,
the $\bar\nu_e$'s propagate like the mass eigenstates owing to the large
matter potential; the $\bar\nu_e$'s do not mix with the other flavor
antineutrinos in this region.
As they propagate to a radius that is larger than $\sim 0.1R_\odot$,
the matter potential becomes smaller; at this stage the $\bar\nu_e$'s
are not the mass eigenstates at all, and mixing with the other flavor
antineutrinos occurs.
This effect induces the decrease at $\agt 0.1R_\odot$ seen in
Fig. \ref{fig:prob_em}.

\subsection{Energy spectrum at the Super-Kamiokande detector
\label{sub:Energy spectrum at the Super-Kamiokande detector}}

With the conversion probabilities given in the previous subsection and
the original neutrino spectrum emitted from the supernova core, we can
calculate the flux of each flavor neutrino.
From this point on, we assume that the distance to the supernova is 10
kpc.
As the original neutrino spectrum, we adopt the results by two groups:
the Lawrence Livermore group \citep{Totani98} and \citet{Thompson02}.
The Livermore spectrum \citep{Totani98} resulted from a calculation
with $20M_\odot$ progenitor models, and we label it LL20.
Recently, that calculation has been criticized, since it lacks relevant
neutrino processes such as neutrino bremsstrahlung and neutrino-nucleon
scattering with nucleon recoils, which were not recognized to be
important at date of calculation.
However, since there are no other successful simulations of a supernova
explosion, it is premature to conclude that their result is no longer
reliable and we adopt their results.
On the other hand, \citet{Thompson02} calculated for three different
mass progenitors $11M_\odot,15M_\odot,$ and $20M_\odot$, and we label
the models TBP11, TBP15, and TBP20, respectively.
Although they did not succeed in simulating the explosion and their data
end at 0.25 s after the core bounce, they included all the relevant
neutrino processes in their calculations.

Using the flux of each flavor neutrino on the Earth and cross sections
of the relevant neutrino interactions at SK, we can calculate the
expected event numbers from future galactic supernova neutrino bursts.
SK is a water \v{C}erenkov detector with 32 ktons of pure water, based
at Kamioka in Japan.
The relevant interactions of neutrinos with water are
\begin{eqnarray}
\bar{\nu}_e + p &\rightarrow& e^+ + n ~\mbox{(CC)},
\label{eq:nuebar+p} \\
\nu_e + e^- &\rightarrow& \nu_e + e^- ~\mbox{(CC and NC)},
\label{eq:nue+e} \\
\bar{\nu}_e + e^- &\rightarrow& \bar{\nu}_e + e^- ~\mbox{(CC and NC)},
\label{eq:nuebar+e} \\
\nu_x + e^- &\rightarrow& \nu_x + e^- ~\mbox{(NC)},
\label{eq:nux+e} \\
\nu_e + \mbox{O} &\rightarrow& e^- + \mbox{F} ~\mbox{(CC)},
\label{eq:nue+O} \\
\bar{\nu}_e + \mbox{O} &\rightarrow& e^+ + \mbox{N} ~\mbox{(CC)},
\label{eq:nuebar+O}
\end{eqnarray}
where CC and NC stand for charged- and neutral-current interactions,
respectively.
SK has restarted observation with lower performance, finishing the
repair of the unfortunate accident which occurred on 12th November
2001.
The effect of the accident on its performance is expected not to be
serious for supernova neutrinos, because the fiducial volume does not
change, and the threshold energy change (from 5 MeV to about 7--8 MeV)
influences the event number very little. 
Although the energy resolution will become about $\sqrt{2}$ times worse,
it does not matter for our considerations.
In the calculations, we used the energy threshold and the energy
resolution after the accident, or in the so-called SK-II phase.
For the cross section of the most relevant interaction
Eq. (\ref{eq:nuebar+p}), we referred to the recent result by
\citet{Strumia03}, and for the other processes we adopted the same ones
as in Ref. \citep{Ando03b}.
As a result, the expected events decreased by $\sim 10$\% in contrast
with the previous calculation in Ref. \citep{Ando03b} under the same
conditions but for the $\bar\nu_ep$ reaction cross section.

\begin{figure}[htbp]
\begin{center}
\includegraphics[width=8cm]{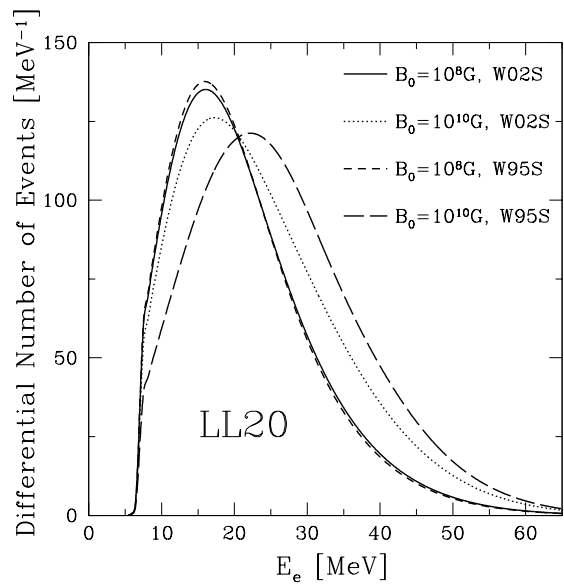}
\caption{Energy spectrum of electrons (positrons) at SK for the first
 0.5 s, obtained using the W02S and W95S presupernova models with
 $B_0=10^8,10^{10}$ G. The original spectrum of LL20 is
 adopted. \label{fig:spectra_SK_Wmodel}}
\end{center}
\end{figure}

\begin{figure}[htbp]
\begin{center}
\includegraphics[width=8cm]{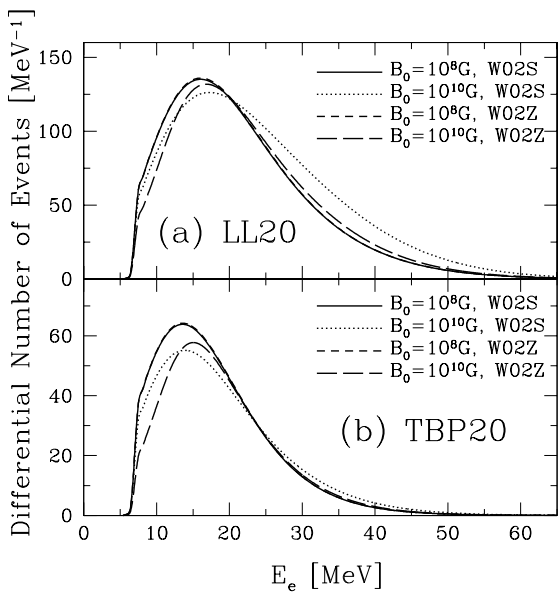}
\caption{Energy spectrum of electrons (positrons) at SK obtained using
 the W02S and W02Z presupernova models with $B_0=10^8,10^{10}$ G. (a)
 The LL20 model for the first 0.5 s. (b) The TBP20 model for the first
 0.25 s. \label{fig:spectra_SK}}
\end{center}
\end{figure}

The expected event number per unit energy range is shown in
Figs. \ref{fig:spectra_SK_Wmodel} and \ref{fig:spectra_SK}.
Figure \ref{fig:spectra_SK_Wmodel} shows the energy spectrum of
electrons (positrons) for the first 0.5 s with the LL20 model, which
was obtained by conversion calculation with the W02S and W95S
presupernova models.
When the magnetic field is not strong, $B_0=10^8$ G, the RSF conversions
are absent, leading to an energy spectrum independent of the
presupernova model.
(The slight difference between the two models comes from the difference
in the adopted mixing angle, i.e., $\tan^2\theta_{12}=0.42$ for W02S and
0.34 for W95S.)
On the other hand, when the magnetic field is sufficiently strong,
$B_0=10^{10}$ G, the energy spectrum is very sensitive to the adopted
presupernova model, reflecting the results of conversion probabilities,
Fig. \ref{fig:prob_em}.
Since the flavor conversions are not as efficient in the case of the
calculations with the W02S model as those with the W95S model, the
expected energy spectrum with the W02S model is not as hard as that with
the W95S model.

Figure \ref{fig:spectra_SK} shows the energy spectrum at SK, obtained
using the W02S and W02Z models with $B_0=10^8,10^{10}$ G.
Figure \ref{fig:spectra_SK}(a) is that for the LL20 model and
Fig. \ref{fig:spectra_SK}(b) is for the TBP20 model; the shapes of the
spectra from the other TBP11 and TBP15 models are almost the same as
that with the TBP20 model, although the absolute value is different.
When $B_0=10^8$ G, because the RSF conversions are absent and pure MSW
flavor transitions occur, the energy spectra of the two models W02S and
W02Z are degenerate.
For the W02Z model with $B_0=10^{10}$ G, since efficient flavor
conversions occur at low energies but not in the high energy region, the
event numbers are suppressed only in the low energy region and high
energy tail is almost the same as that with $B_0=10^8$ G.
Since the average energy difference between the $\bar\nu_e$'s and
$\nu_{\mu,\tau}$'s is not prominent for the TBP models, the difference
between energy spectra with various presupernova models is suppressed,
compared with the LL20 model, particularly in the high energy tail.

\section{DISCUSSION \label{sec:DISCUSSION}}

\subsection{What can we learn about the RSF effect from the neutrino
  signal? \label{sub:What can we learn about the RSF effect from the
  neutrino signal?}}

In the previous section, it was shown that the RSF conversion strongly
depends on the presupernova models of various metallicities.
The behavior of the flavor mixing in the supernova envelope is very
different from one presupernova model to another, and they indicate
various profiles in supernovae such as magnetic field strength and the
value of $(1-2Y_e)$ at the resonance points as well as the neutrino
magnetic moment.
Unfortunately, however, what we can observe is the energy spectrum of
electrons (positrons) at detectors on the Earth alone, and much
information on the detailed character is lost, and therefore we can
obtain only rather rough characteristics in principle.
Here, we use as a simple indicator of the RSF conversions the following
quantity:
\begin{equation}
R_{\rm SK}=\frac{\mbox{number of events for } E_e>25~{\rm MeV}}
 {\mbox{number of events for } E_e<20~{\rm MeV}}.
\label{eq:R_SK}
\end{equation}
The values of $R_{\rm SK}$ for various models are summarized in Table
\ref{table:R_SK}.
Even if we use the data for the first 0.25 or 0.5 s, it appears that
statistically sufficient discussions are possible.

\begin{table}[htbp]
\begin{center}
\caption{The values of $R_{\rm SK}$ for various models. Attached errors
 are statistical ones at the 1$\sigma$ level. \label{table:R_SK}}
\begin{tabular}{lcccc}\hline\hline
 & \multicolumn{2}{c}{$B_0=10^8$ G}
 & \multicolumn{2}{c}{$B_0=10^{10}$ G} \\
 Model & W02S & W02Z & W02S & W02Z \\ \hline
 LL20 (0.5 s) & $0.59\pm 0.02$ & $0.59\pm 0.02$
 & $0.96\pm 0.04$ & $0.72\pm 0.03$\\
 LL20 (0.25 s) & $0.41\pm 0.03$ & $0.41\pm 0.03$
 & $0.71\pm 0.04$ & $0.52\pm 0.03$ \\
 TBP20 & $0.25\pm 0.02$ & $0.25\pm 0.02$
 & $0.35\pm 0.03$ & $0.32\pm 0.03$ \\
 TBP15 & $0.21\pm 0.03$ & $0.21\pm 0.03$
 & $0.33\pm 0.04$ & $0.27\pm 0.03$\\
 TBP11 & $0.22\pm 0.03$ & $0.22\pm 0.03$
 & $0.35\pm 0.03$ & $0.28\pm 0.03$\\ \hline\hline
\end{tabular}
\end{center}
\end{table}

In practice, to make matters worse, there is a large uncertainty
concerning the original neutrino spectrum emitted by the core collapse;
actually the values of $R_{\rm SK}$ are very different between the LL20
and TBP20 models (see Table \ref{table:R_SK}).
Thus, we must reduce the systematic errors of the models in discussing
the RSF effect from the energy spectrum obtained.
At present, however, this problem is very difficult and there is no
way but to wait for the future development of numerical simulation of
supernova explosions.
Therefore, at present, it is very difficult to say even whether the RSF
effect actually occurred or not.

If the systematic errors concerning the original neutrino spectrum are
considerably reduced by future development of numerical simulations,
it is expected that useful implications for the RSF mechanism will be
obtained from the value of $R_{\rm SK}$.
Therefore, at the end of this subsection, we discuss to what extent we
can learn from the observed neutrino signal {\it assuming} that the
systematic errors are much reduced.
If the magnetic field or the neutrino magnetic moment are too small to
induce adiabatic RSF conversions, the metallicity of presupernova
stars is not detectable as shown in the second and third columns of Table
\ref{table:R_SK}.
On the other hand, in the case of a strong magnetic field like
$B_0=10^{10}$ G, the value of $R_{\rm SK}$ becomes larger than that for
$B_0=10^8$ G at the $\agt (2$--$3)\sigma$ level, which is a statistically
meaningful number.
However, it is very difficult to estimate the metallicity of the
presupernova star, because the value of $R_{\rm SK}$ sensitively depends
on the value of $B_0$, when it is larger than $10^9$ G as shown in
Ref. \citep{Ando03b}.
Consequently, even if the systematic errors were considerably reduced in
the future, all we could say concerning the RSF effect from the galactic
supernova neutrino burst would be at most that the RSF conversions have
occurred or not.
Other detailed discussion concerning various quantities such as the
magnetic field strength, supernova metallicity, and neutrino magnetic
moment would not be accessible.

\subsection{Future prospects \label{sub:Future prospects}}

In the near future, it is expected that the many uncertainties which
complicate the observations of the RSF conversion will be much reduced.
These uncertainties are, e.g., the neutrino magnetic moment, the
supernova magnetic field structure, and the metallicity of the
supernova.

If KamLAND receives a positive signal for the appearance of solar
$\bar\nu_e$, then it indicates that the spin-flavor conversion of the
{\it Majorana} neutrino actually occurs inside the Sun, and from the
data we can obtain implications for the nonzero value of the
neutrino magnetic moment.
Actually, in the Sun the relevant RSF conversion is $\nu_e
\leftrightarrow\bar\nu_{\mu,\tau}$, whereas in supernovae it is
$\bar\nu_e\leftrightarrow\nu_{\mu,\tau}$.
However, if the neutrinos are Majorana particles, the absolute values of
the transition magnetic moments, which are responsible for the above two
processes, are the same (the sign is different; $\mu_{ij}=-\mu_{ji}$).
Thus, depending on the value that is observed by the solar neutrinos,
it can easily be estimated whether the RSF effect is the relevant
conversion process for the supernova neutrinos; further, if it is the
relevant process, the magnetic field strength itself, not the
combination $\mu_\nu B$, at the resonance points can be inferred.
From future KamLAND results we can constrain only one of the magnetic
moment tensor elements, which consist of three independent quantities,
but we need all three values for the supernova case.
Yet that information, if actually obtained, would be helpful for the
estimation of other values and very useful.

As for the metallicity, we will obtain useful inferences from
astrophysical discussions.
The lifetime of massive stars which end their life by gravitational
collapse is much shorter than that of the Sun, and the progenitors of
observed supernovae are, therefore, younger.
Consequently, the metallicity of the galactic supernovae is expected to
be at least the solar abundance or more metal rich.
If the metallicity is higher than that of the Sun, the suppression of
$(1-2Y_e)$ will be weaker and the RSF conversion will incline to be more
adiabatic.
On the other hand, the large and small Magellanic clouds are known to be
very metal-poor systems \citep{Hilker95,Meliani95}.
Thus, the determination of the explosion site will provide several rough
estimations of the metallicity.
Even though the explosion occurred in an optically thick environment
such as the galactic center, the detected neutrinos alone could locate
the supernova precisely; from the event number the distance to a
supernova will be inferred, and from the event distribution the
supernova direction can be determined \citep{Ando02}.

Another object that is related to the metallicity effect is the
supernova relic neutrino (SRN) \citep{Ando03a,Ando03c}.
Because the SRN is the accumulation of neutrinos from all the past
supernovae, the SRN includes neutrinos from supernovae with quite low
metallicity in the early phase of galaxy formation.
Recently, the SK collaboration released the first result of their search
for the SRN signal \citep{Malek03}, and it is only about a factor of 3
larger than the theoretical predictions by \citet{Ando03a}.
Data accumulation for several more years may be decisive for actual
detection and setting a severe constraint on the SRN flux.
However, because the totally time-integrated neutrino spectrum during
the neutrino burst is needed for the SRN calculation, it is very
difficulty to estimate the SRN flux including the RSF effect, which is
affected by the time-dependent density and magnetic field profiles.
Thus, it will be difficult to derive useful information about the RSF
conversions from future SRN observations.

\section{CONCLUSION \label{CONCLUSION}}

The RSF conversion of supernova neutrinos was investigated in detail;
this conversion is induced by a nonzero magnetic moment of neutrinos and
a strong supernova magnetic field.
Because of the effective potential for the RSF conversion which
delicately depends on the deviation of $Y_e$ from 0.5, we studied the
dependence of the RSF conversion on the adopted presupernova models.
As the models, we used recent models by \citet{Woosley02} with
solar and zero metallicities (W02S and W02Z) as well as that by
\citet{Woosley95} with solar metallicity (W95S), which was the only
model adopted in our previous paper \citep{Ando03b}, for comparison.

The W02 models were calculated using a recent shell model and
resulted in substantial revisions to the older data sets in the W95S
model.
As a result, for both W02 models the value of $(1-2Y_e)$ suddenly
drops at the radius where the RSF conversions occur, leading to a
less efficient RSF conversion than that with the W95S model, whose
$(1-2Y_e)$ gradually decreases at the RSF regions.
In particular, for the W02Z model, the value of $(1-2Y_e)$ becomes
exactly zero at the boundary between the Si + O and O + Ne + Mg layers,
which makes the conversion almost completely nonadiabatic.
Thus, we found that energy spectra expected at SK for the W02 models
would not be as hard as those for the W95S model.
Unfortunately, however, there are many uncertainties at present, which
are concerned with, e.g., the original neutrino spectrum, the neutrino
magnetic moment, the supernova magnetic field structure, and the
supernova metallicity.
All these uncertainties complicate the discussions, and it is very
difficult to obtain some useful information on the supernova metallicity
or magnetic field strength, because the detected energy spectrum does
not indicate which effect contributes to what extent.

In the future, however, it is expected that the development of
numerical simulations of supernova explosions will much reduce the
uncertainties for the original neutrino spectrum, and the KamLAND solar
$\bar\nu_e$ observation, whether detected or not, will provide useful
information on the neutrino magnetic moment.
We believe that all these developments will enable a more decisive
investigation of the RSF conversion of supernova neutrinos.

\begin{acknowledgments}
S. A.'s work is supported by Grants-in-Aid for JSPS Fellows.
K. S.'s work is supported in part by Grants-in-Aid for scientific research
 provided by the Ministry of Education, Science and Culture of Japan
 through Research Grant No. S14102004 and No. S14079202.
\end{acknowledgments}

\bibliography{refs}

\end{document}